# Influence of salt on membrane rigidity of neutral DOPC vesicles

Judith U. De Mel,[1] Sudipta Gupta,[1*] Rasangi M. Perera,[1] Ly Ngo,[1] Piotr Zolnierczuk,[2] Markus Bleuel,[3] Sai Venkatesh Pingali,[4] and Gerald J. Schneider[1,5*]

*[1]Department of Chemistry, Louisiana State University, Baton Rouge, LA 70803, USA*

*[2]Jülich Centre for Neutron Science (JCNS), Outstation at SNS, POB 2008, 1 Bethel Valley Road, TN 37831, Oak Ridge, USA*

*[3]NIST Center for Neutron Research, National Institute of Standards and Technology, Gaithersburg, Maryland 20899-8562, USA*

*[4]Biology and Soft Matter Division, Neutron Sciences Directorate, Oak Ridge National Laboratory (ORNL), POB 2008, 1 Bethel Valley Road, TN 37831, Oak Ridge, USA*

*[5]Department of Physics & Astronomy, Louisiana State University, Baton Rouge, LA 70803, USA*

# 1 Abstract

Salt is a very common molecule in aqueous environments but the question of whether the interactions of monovalent ions $Na^+$ and $Cl^-$ ,with the neutral heads of phospholipids are impactful enough to change the membrane rigidity is still a mystery. To provide a resolution to this long simmering debate, we investigated the dynamics of DOPC vesicles in the fluid phase with increasing external salt concentration. At higher salt concentrations, we observe an increase in bending rigidity from neutron spin echo spectroscopy (NSE) and an increase in bilayer thickness from small-angle X-ray scattering (SAXS). We compared different models to distinguish membrane undulations, lipid tail motions and the translational diffusion of the vesicles. All the models indicate an increase in bending rigidity by a factor of 1.3 to 3.6. We demonstrate that even for $t > 10$ ns, and for $Q > 0.07$ Å $^{-1}$ the observed NSE relaxation spectra is clearly influenced by the translational diffusion of the vesicles. For $t < 5$ ns, the lipid tail motions dominate the intermediate dynamic structure factor. As the salt concentration increases this contribution diminishes. We introduced a new time-dependent analysis for the bending rigidity that highlights only a limited Zilman-Granek time window where the rigidity is physically meaningful.

TOC Figure

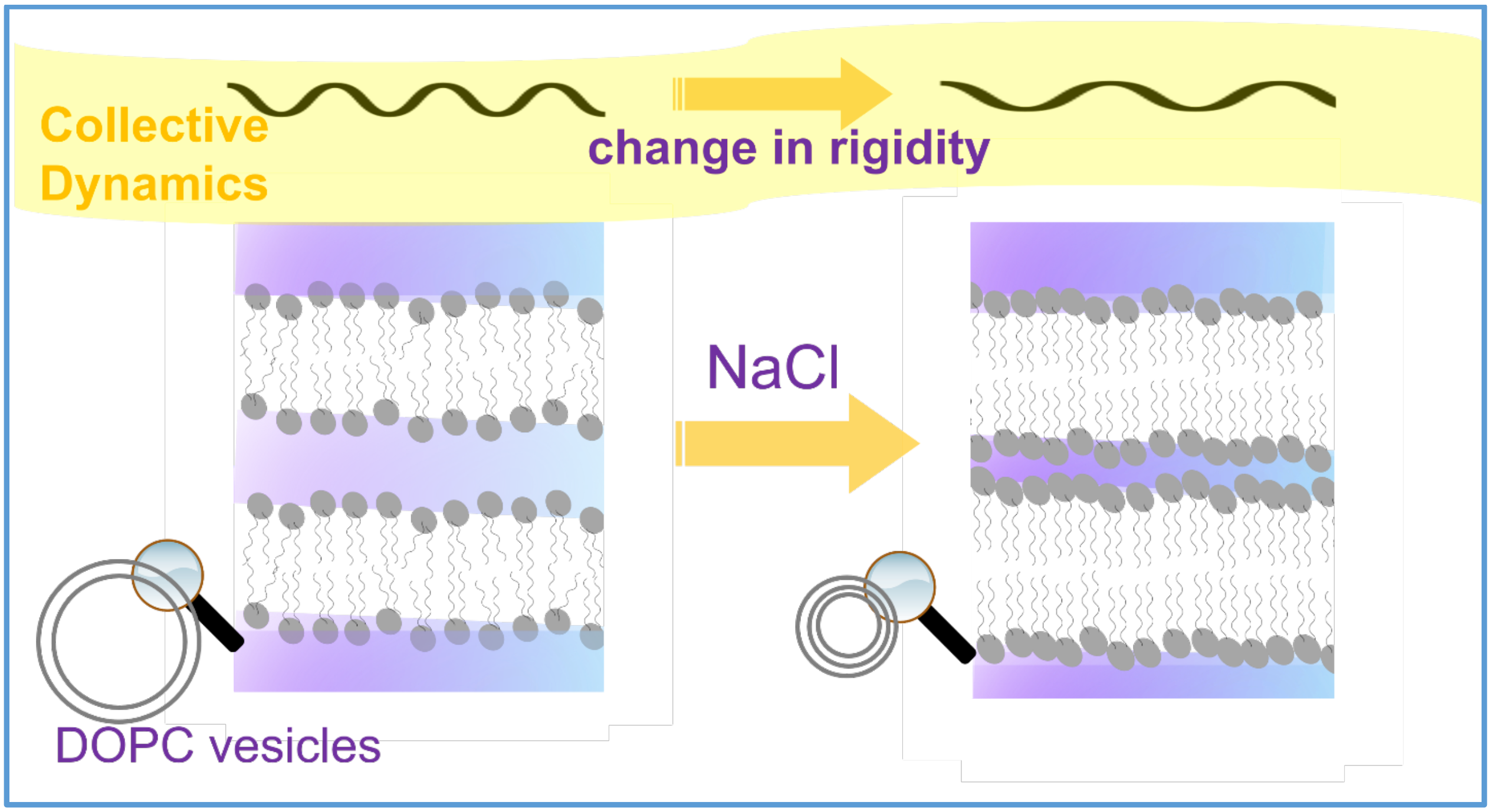

# 2 INTRODUCTION

The aqueous environment surrounding biological membranes contains many ions, including $Na^{+}$, $K^{+}$, $Mg^{2+}$ and $Ca^{2+}$ and $Cl^{-}$. The interactions among ions and cell membrane are assumed to trigger multiple significant physiological activities including membrane fusion, phase transformations, and regulation of ion channels located at membranes.[1-4] Several studies indicate a substantial impact of ions on the morphology and function of lipid bilayers which are an essential part of living cells.[5, 6]

The influence of NaCl and other monovalent salts on liposome size and bilayer thickness of phosphotadylcholine has been explored by a variety of tools for many years. Despite the many studies conducted, there are contradictions in the explanations and a coherent picture is lacking. For example, several studies have reported changes in the size of vesicles with the addition of monovalent salts. Claessens *et.al*. report size decrease of 1,2-di-(octadecenoyl)-*sn*-glycero-3-phosphocholine (DOPC) vesicles above 50 mM in NaBr up to 150 mM and an increase again up to 400 mM.[7] In a 2005 study, Yue and colleagues show that unilamellar vesicles (ULVs) from 1,2-dimyristoyl-sn-glycero-3-phosphocholine (DMPC) can change the size upon NaCl addition using dynamic light scattering and small-angle neutron scattering. Their results are contradicting to that of Claessens, which shows an overall increase in the size. It is also noteworthy that they report no size change occurs if the vesicles were doped in NaCl.[8]

A crucial property of phospholipid membranes, especially vesicles, is their membrane rigidity. It is extremely important to understand how the membrane rigidity changes or can be manipulated in the presence of NaCl in order to improve nano-carrier interactions with cell membranes, improved transportation strategies, and other nanomaterial applications using phospholipid vesicles. Most liposomal formulations are prepared in saline solutions or at a certain pH controlled with an ionic strength that's physiologically relevant. Therefore, it is important from biophysical perspective and beyond to understand how vesicles change membrane rigidity in such systems. In existing literature, there are several discrepancies when we consider membrane rigidity changes induced by monovalent salts. While some experiments lead to the conclusion that monovalent salts have no significant impact on membrane rigidity, others have reported an increase or decrease of rigidity. X-ray diffraction studies on 1,2-dilinoleoyl-sn-glycero-3-phosphocholine (DLPC) multilamellar membranes have shown no significant changes in membrane rigidity in the presence of KCl or KBr.[9] However, the same study has reported about a 50% decrease in van der Waals strength in the presence of salt. Pabst *et.al.* report there is no significant membrane rigidity change up to 1000 mM of a monovalent salt concentration and increase of membrane rigidity beyond this limit by utilizing small-angle x-ray diffraction on POPC membranes.[10] Claessens *et.al.* report an increasing bending rigidity for zwitterionic DOPC membranes with the increasing ionic strength.[7] Another study conducted on phosphocholine Giant Unilamellar Vesicles (GUVs) has revealed that 100 mM NaCl can decrease bending rigidity among effects from other salt mixtures and buffer solutions.[11] Tenchov *et.al.* synthesized LUVs using lipids extracted from salt intolerant or moderately tolerant microorganisms and measured the leakage of entrapped fluorescent dye in 100-400 mM NaCl range where they observed rapid leakage of fluorescent dye in comparison to the LUVs prepared by salt tolerant microorganism lipids.[12] Very recently Faizi *et.al.* conducted experiments on GUVs varying the fraction of lipids with charged head groups in the presence of NaCl. They report a rapid increase in membrane rigidity with surface charge and a reduction of the same when exposed to NaCl.[13] Evidently, the question about membrane rigidity in the presence of NaCl is unresolved and we will be exploring that in-depth in this study.

In this paper we use a holistic approach to shed more light on the impact of NaCl on DOPC vesicles. We focus on NaCl introduced externally to the vesicles in a concentration ranging from 0 – 500 mM, which is more relevant since $Na^+$ dominates in the extracellular media of the cells in the concentration range of roughly 100 – 150 mM pairing with $Cl^-$ as the counter ions. The use of phosphocholine vesicles is especially relevant for studying interactions with external lipid bilayer of the mammalian cell membrane since their presence is predominant in the outer lipid layer in comparison to the inner layer.[14, 15] For that purpose, we use a series of techniques such as cryogenic transmission electron microscopy (cryo-TEM), small-angle X-ray scattering (SAXS), small-angle neutron scattering (SANS), viscometry for complimentary structural characterizations followed by neutron spin echo (NSE) spectroscopy to examine the collective lipid dynamics leading to rigidity.

# 3 EXPERIMENTAL SECTION

## 3.1 Materials

All chemicals and reagents were used as received. 1,2-di-(octadecenoyl)-*sn*-glycero-3-phosphocholine (DOPC) was purchased from Avanti Polar Lipids (Alabaster, AL, USA), biotechnology grade Sodium Chloride (NaCl) (99.9% purity) was obtained from VWR Life Sciences (Solon, OH, USA), organic solvents (HPLC grade) and $D_2O$ were received from Sigma Aldrich (St. Louis, MO, USA).

## 3.2 Sample preparation

DOPC vesicles were prepared by dissolving DOPC lipid powder in chloroform and removing the solvent using a rotary evaporator and drying further under vacuum overnight. The dried lipid was hydrated using $D_2O$ and the resultant solution was subjected to freeze-thaw cycling by alternatingly immersing the flask in water at around 50 °C and placing in a freezer at -20 °C in ten minute intervals. Finally, the solution was extruded using a mini extruder (Avanti Polar Lipids, Alabaster, AL, USA) through a polycarbonate membrane with pore diameter of 100 nm (33 passes) to obtain unilamellar vesicles. Vesicle solutions were mixed with NaCl solutions to obtain the desired external ionic concentrations. In this context any osmotic effect will be a subject of future publications. Measurements for each mixture were averaged starting 24 hours after sample preparation. All experiments were conducted at ambient temperature, 23 °C.

## 3.3 Characterization

**Small-Angle Neutron Scattering (SANS) and Very small-angle neutron scattering (VSANS)** Small-angle neutron scattering (SANS) studies were performed at the CG-3 Bio-SANS instrument at the High-Flux Isotope Reactor (HFIR) of Oak Ridge National Laboratory.[16] The sample-to-detector distance was fixed to 15.5 m, at neutron wavelength, $\lambda$ = 6 Å. This configuration covers a $Q$ - range from 0.003 to 0.6 Å$^{-1}$, where $Q = 4\pi \sin(\theta/2)/\lambda$, with the scattering angle, $\theta$. A wavelength resolution of, $\Delta\lambda/\lambda$ = 13%, was used. A typical SANS data reduction protocol, which consisted of subtracting scattering contributions from the empty cell (2 mm, Hellma cells) and background scattering was used to yield absolute calibrated intensities, $I(Q)$. Data reduction was conducted employing the Mantid plot software. Very-small-angle neutron scattering (VSANS) studies were performed at the VSANS instrument at the NIST Center for Neutron

Research. The samples were measured with two instrument configurations, resulting in a collimation length of 24 and 10 m, respectively. Two sample detector distances, 4 m and 19 m, for the zero guide configuration, and 2 m and 9 m for the 7 guide configuration were utilized. The neutron wavelength was $\lambda$ = 6 Å. These configurations cover a $Q$ - range from 0.0018 to 0.11 Å$^{-1}$ and from 0.0067 to 0.34 Å$^{-1}$.

**Small-Angle X-ray Scattering (SAXS)** Small-angle X-ray (SAXS) scattering experiments were conducted at LIX at the National Synchrotron Light Source II, Brookhaven National Laboratory and at a SAXSpace (Anton Paar) instrument with a micro-focus X-ray source ($\lambda$ = 0.154 nm) at Oak Ridge National Laboratory (ORNL). At the synchrotron instrument, the samples were measured in a flow cell with an acquisition time of 1 s, whereas the samples were loaded in 1 mm borosilicate glass capillary cylinders for the lab X-ray with an acquisition time of 10 s. The recorded intensities were corrected for dark current, empty cell and solvent (buffer) using standard procedures.[17] [18] The scattering intensity was normalized to absolute units (cm$^{-1}$) using water as calibration standard.[19]

**Cryo-Transmission Electron Microscopy (Cryo-TEM)** Cryo-transmission electron microscopy (TEM) images were recorded on a Tecnai G2 F30 operated at 150 kV. A volume of ten microliters of the sample (0.125 wt% DOPC: in pure $D_2O$, 150 and 500 mM external NaCl) was applied to a 200 mesh lacey carbon grid mounted on the plunging station of an FEI Vitrobot™ and excess liquid was blotted for 2 s by the filter paper attached to the arms of the Vitrobot. The carbon grids with the attached thin film of vesicle suspensions were plunged in to liquid ethane and transferred to a single tilt cryo - specimen holder for imaging. Cryo-TEM images were obtained in the bright field setting. Since cryo-TEM can only be conducted at dilute vesicle concentrations, the concentration of vesicles is different to other listed techniques and therefore is strictly used for visualization without any statistical analysis.

**Neutron Spin Echo (NSE) Spectroscopy** We collected NSE data at BL15 at the Spallation Neutron Source of the Oak Ridge National Laboratory, Oak Ridge, TN[20] and at the NGA-NSE at the NIST Center for Neutron Research (NCNR) of the National Institute of Standards and Technology (NIST).[21] We used Hellma quartz cells at BL15-NSE and Titanium cells at NGA-NSE, in both cases with 4 mm sample thickness. Lipid concentration was always 5%. The data reduction was performed with the standard ECHODET software package of the SNS-NSE instrument and Dave[22] software package for NGA-NSE. Wavelengths of 8, 11 and 15 Å were used at NGA-NSE and 8 Å was used at BL15-NSE. $D_2O$ was measured separately and subtracted as background.

# 4 THEORETICAL BACKGROUND

## 4.1 Structure

Hereafter the derivation of the macroscopic scattering cross-section, $d\Sigma/d\Omega$, for the bilayer and the vesicle structure is presented. We use the fact that the SAXS and SANS experiments were conducted at ambient temperature, 23 °C, which corresponds to the fluid phase of DOPC.[23]

**Bilayer structure**: The random lamellar sheet consisting of the heads and tails of the phospholipids can be modelled using the Caille structure factor.[24, 25] It provides direct access to the macroscopic scattering cross-section given by the scattering intensity for a random distribution of the lamellar phase, as:

$$\frac{d\Sigma}{d\Omega}(Q)_{SAXS} = 2\pi \frac{VP(Q)S(Q)}{Q^2 d} \tag{1}$$

with the scattering volume, $V$, and the lamellar repeat distance, $d$. The form factor is given by:

$$P(Q) = \frac{4}{Q^2}\left[\Delta\rho_{\mathrm{H}}\left\{\sin\left(Q(\delta_H+\delta_T)\right) - \sin(Q\delta_T)\right\} + \Delta\rho_{\mathrm{T}}\sin(Q\delta_T)\right]^2 \tag{2}$$

The scattering contrasts for the head and tail are $\Delta\rho_{\mathrm{H}}$ and $\Delta\rho_{\mathrm{T}}$, respectively. The corresponding thicknesses are $\delta_H$ and $\delta_T$, respectively, as presented in Figure 1. The head to head bilayer thickness is given by, $\delta_{HH} = 2(\delta_H+\delta_T)$. The Caille structure factor is given by

$$S(Q) = 1 + 2\sum_{n=1}^{N-1}\left(1-\frac{n}{N}\right)\cos(Qdn)\exp\left(-\frac{2Q^2 d^2 \alpha(n)}{2}\right) \tag{3}$$

with the number of lamellar plates, $N$, and the correlation function for the lamellae, $\alpha(n)$, defined by

$$\alpha(n) = \frac{\eta_{\rm cp}}{4\pi^2}(\ln(\pi n) + \gamma_E) \tag{4}$$

with $\gamma_E$ = 0.57721 the Euler's constant. The elastic constant for the membranes are expressed in terms of the Caille parameter, $\eta_{\rm cp} = \frac{{Q_1}^2 k_B T}{8\pi\sqrt{\kappa_c \bar{B}}}$, where $\kappa_c$ and $\bar{B}$ are the bending elasticity and the compression modulus of the membranes. Here $\bar{B}$ is associated with the interactions between the membranes. The position of the first-order Bragg peak is given by $Q_1$, whereas, $k_B$ is the Boltzmann's constant and $T$ the absolute temperature. Each of the lengths, such as, $d$, $\delta_H$ and $\delta_T$ are convoluted with a Gaussian distribution function to account for the thickness polydispersity.

**Vesicle structure:** The vesicle form factor is modeled using an extension of the core-shell model used in our previous studies.[26, 27] The core is filled with water and in case of multilamellar vesicles encapsulated by $N$ shells of lipids and $N$-1 layers of solvent as illustrated in *Figure 1*. Each shell thickness and scattering length density is assumed to be constant for the respective shell. The 1D scattering pattern is described by:

$$P(Q, R, t, \Delta\rho) = \frac{\phi[F(Q)]^2}{V(R_N)} \tag{5}$$

with

$$F(Q) = (\rho_{\rm shell} - \rho_{\rm solv}) \sum_{i=1}^{N} \left[ 3V(r_i) \frac{\sin(Qr_i) - Qr_i \cos(Qr_i)}{(Qr_i)^3} - 3V(R_i) \frac{\sin(QR_i) - QR_i \cos(QR_i)}{(QR_i)^3} \right] \tag{6}$$

For

$$r_i = r_c + (i-1)(t_s + t_w) \quad \text{solvent radius before shell } i$$
$$R_i = r_i + t_s \quad \text{shell radius for shell } i \tag{7}$$

Here, *V(r)* is volume of the sphere with radius *r*, $r_c$ is the radius of the core, $t_s$ is the thickness of the individual shells, $t_w$ is the thickness of the interleaved solvent layers, ϕ, the corresponding lipid volume fraction. For DOPC we used the neutron scattering length density (NSLD) of the shell, $\rho_{\text{shell}} = 3.01 \times 10^9$ cm$^{-2}$ and for $D_2O$ the NSLD of the solvent, $\rho_{\text{solv}} = 6.36 \times 10^{10}$ cm$^{-2}$, respectively.[28] The macroscopic scattering cross-section is obtained by

$$\frac{d\Sigma}{d\Omega}(Q)_{SANS} = \int drP(Q, R, t, \rho_{\text{lipo}}, \rho_{\text{solv}})s(r) \tag{8}$$

For the size polydispersity, $s(r)$, we used a Schulz distribution and a log-normal distribution.[18] In addition, the thickness of the shell and the solvent are convoluted with a Gaussian distribution function to account for the thickness polydispersity.

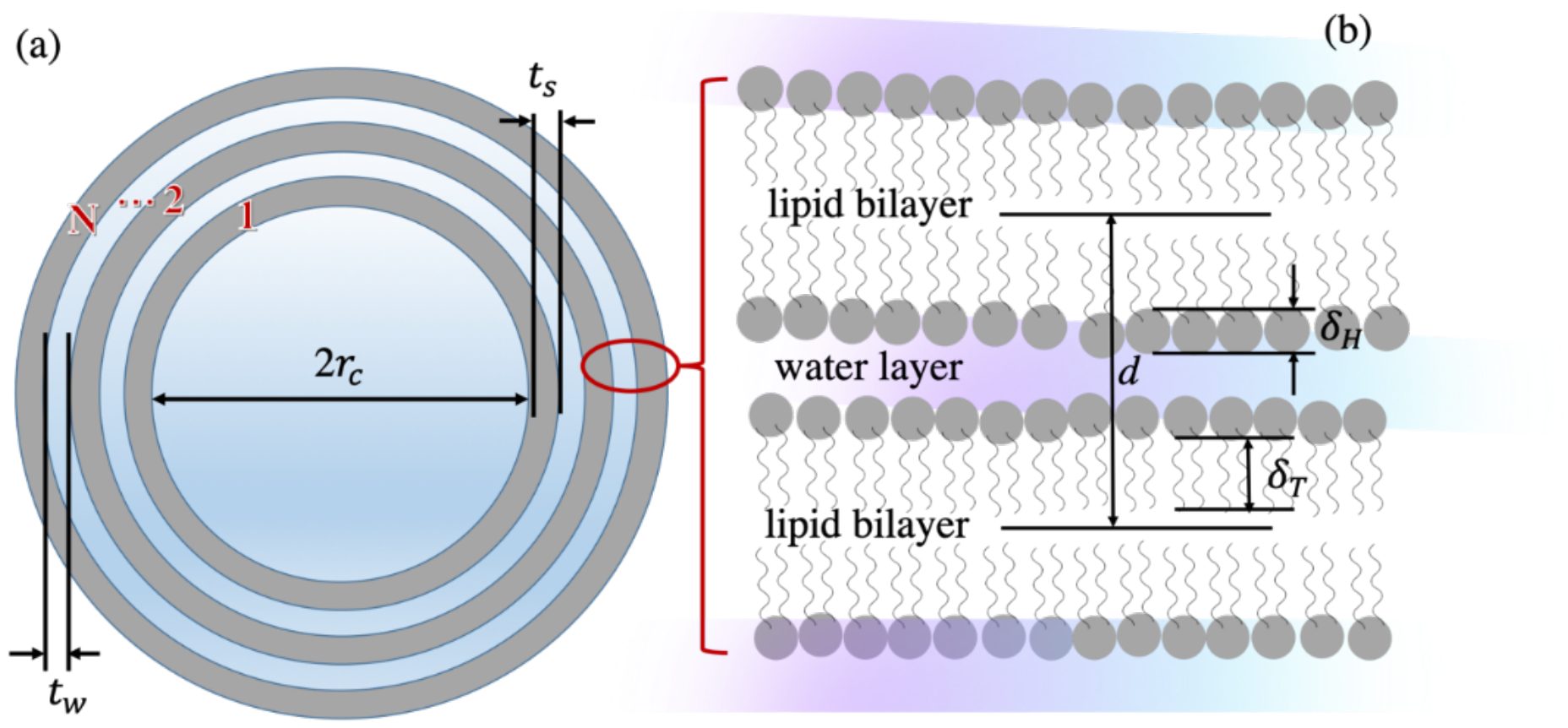

*Figure 1. Schematic representation of (a) the multilamellar vesicle, and (b) lipid multilayers illustrating the number of bilayers, N, the radius of the core,* $r_c$*, the thickness of the individual shells,* $t_s$*, the thickness of the interleaved solvent layers,* $t_w$*, the thickness of the lipid head,* $\delta_H$*, the thickness of the lipid tail region,* $\delta_T$ *and the lamellar repeat distance, d, of bilayers.*

## 4.2 Dynamics of the lipid bilayer

Neutron spin echo (NSE) spectroscopy has proven to be a powerful tool to follow the molecular motions in vesicles.[28-30] This method reaches the highest energy resolution (~ neV) of all available neutron scattering spectrometers and therefore allows to measure the dynamic structure factor or the intermediate scattering function, *S*(*Q*,*t*), up to several hundred nanoseconds.

Recently, it has been shown that diffusion, membrane fluctuations, and confined motion of lipid tails lead to major contributions for the modeling of the intermediate scattering function.[31] The statistically independent tail-motion and height-height correlation resulting in membrane undulations can be coupled to the overall translational diffusion of the liposome as: [31]

$$S_{liposome}(Q,t) = \left( n_{H,head} + n_{H,tail}\left(\mathcal{A}(Q) + \left(1-\mathcal{A}(Q)\right)\exp\left(-\left(\frac{t}{\tau}\right)^{\beta}\right)\right)\right) S_{ZG}(Q,t)\exp(-D_t Q^2 t) \tag{9}$$

Here the relative fractions of protons in the head is kept fixed to, $n_{H,head}$ = 0.21, for h-DOPC, and, $n_{H,tail} = 1 - n_{H,head}$ = 0.79. More precisely the number of protons in this case reflects the average scattering length density of the head or tail of the lipid. The first part considers the motion of the lipid molecule, whereas the second and third part considers the undulations, which can be well described by the Zilman-Granek (ZG) approach given by a stretched-exponential decay: [32]

$$S_{ZG}(Q,t) \propto \exp\left[-\left(\Gamma_Q t\right)^{2/3}\right]. \tag{10}$$

The only free parameter is the *Q*-dependent decay rate, $\Gamma_Q$, from which we derive the intrinsic bending modulus, $\kappa_\eta$, given by[28, 33, 34]

$$\Gamma_q = 0.0069\gamma\frac{k_B T}{\eta}\sqrt{\frac{k_B T}{\kappa_\eta}} \tag{11}$$

Here $\eta$ is the viscosity, $k_B$ the Boltzmann constant, *T* the temperature, and $\gamma$ is a weak, monotonously increasing function of $\kappa_\eta/k_B T$.[32] In case of lipid bilayers usually $\kappa_\eta/k_B T \gg 1$, leading to $\gamma$= 1.[28, 29, 32, 34, 35] More details can be found in a recent review.[36]

Additionally, we analyze the mean squared displacement ($\langle \Delta r(t)^2 \rangle$, MSD) and the non-Gaussianity parameter, $\alpha_2(t) = \frac{d}{d+2}\frac{\langle \Delta r(t)^4\rangle}{\langle \Delta r(t)^2\rangle^2} - 1$, from the measured dynamic structure factor, *S*(*Q*,*t*), using a cumulant expansion given by, [27, 31, 37, 38]

$$\frac{S(Q,t)}{S(Q)} = A \exp\left[-\frac{Q^2\langle\Delta r(t)^2\rangle}{6} + \frac{Q^4\alpha_2(t)}{72}\langle\Delta r(t)^2\rangle^2\right] \tag{12}$$

The non-Gaussianity parameter $\alpha_2$ is essentially defined as quotient of the fourth $\langle\Delta r(t)^4\rangle$ and the second moment squared $\langle\Delta r(t)^2\rangle^2$ and $d = 3$, is the dimension of space.[27, 38, 39] Following equation 10 and 11 one can express the membrane rigidity as a function of Fourier time, given by[31]

$$\frac{\kappa_\eta}{k_B T} = \frac{t^2}{c(\eta,T)^3\langle\Delta r(t)^2\rangle^3} \tag{13}$$

with $c(\eta,T) = \frac{1}{6}\left(\frac{\eta}{0.0069 k_B T}\right)^{2/3}$. For ZG approximation $\langle\Delta r(t)^2\rangle \propto t^{2/3}$ and the bending rigidity as a function of time should yield, $\kappa_\eta/k_B T \propto t^2/t^2$ = constant. Any deviation from this constant behavior will reflect additional dynamics that is not taken into account by the ZG model.

## 5 RESULTS AND DISCUSSION

Hereafter we introduce our results on the membrane dynamics by NSE. Structural parameters are deducted from SAXS and SANS. We use SANS results to determine the diameter of the vesicles and SAXS to obtain information on the individual layers. The advantage of X-rays is the good contrast of the phosphoric head groups of the lipids, and the excellent resolution, $\Delta Q/Q$, is negligible compared to the finite width of the structure peaks in liposomes.

In order to maintain the same conditions for SAXS/SANS and NSE, all experiments use 5 wt% of DOPC. Concentrations of 5 wt% and above are very common for NSE experiments,[27, 34, 36] as viable compromise between good statistics and measurement time. All measurements have been conducted at $T$ = 20 °C (fluid phase of DOPC)[36] for three NaCl concentrations, 0, 150 and 470 mM. The phase transition temperature for DOPC lipid is very low, $T_m$ = -16.5 °C,[23] so we neglect any small contributions due to phase transition in all our samples.

### 5.1 Morphology by SAXS and SANS

We begin with the determination of the diameter of vesicles from SANS and continue with the analysis of the bilayers from SAXS. Figure 2 (a) illustrates SANS data for different NaCl concentrations. The data is vertically scaled for the convenience of the reader. We observe a decreasing intensity with increasing the momentum transfer, Q. With the addition of salt, peaks emerge at high $Q$ and become more pronounced with increasing salt concentration. At 0 mM there is a slight peak which is hardly visible. Solid lines represent the modeling as described in equations 5-8.

We now need to ask how these SANS diagrams are connected to the diameter of the vesicles. Since the neutron scattering cross section of protons is rather large, SANS contrast on vesicles essentially results from protons. Most protons in protonated DOPCs are in the tail region. The bilayer thickness determined by SANS belongs essentially to the thickness defined by the hydrophobic core.[24] The thickness of the individual shells, $t_s$ , from SANS is smaller than that of the head-to-head bilayer thickness, $\delta_{HH}$, SAXS. The corresponding outer perimeter radius is given by $R_{SANS}$, and an estimated number of lipids per vesicle or the aggregation number is given by $N_{agg}$. We observe a clear increase in $N_{agg}$ at higher concentration.

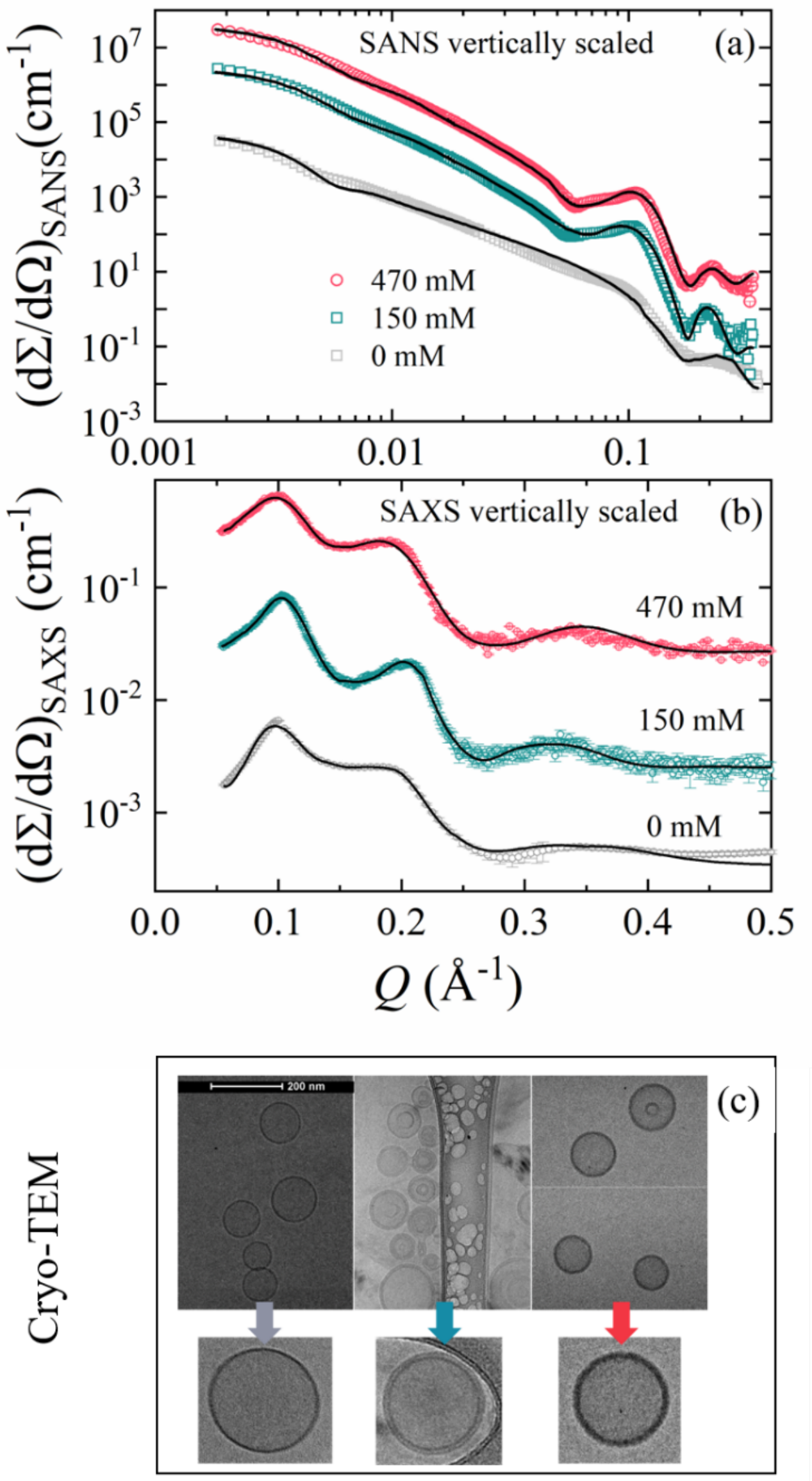

*Figure 2. (a) SANS scattering intensity for various NaCl concentrations added from outside on DOPC dispersed in $D_2O$. The solid lines represent fits using the model introduced by equation 5-8. (b) Corresponding SAXS data for various NaCl concentrations. SAXS and SANS data is vertically scaled for better visualization by multiplication with a constant value. The solid lines are the model as described in equations 1-4. (c) Cryo-TEM images of DOPC vesicles in 0, 150 and 470 mM (left to right) NaCl concentrations.*

The evolution of the peak at Q ≈ 0.1 Å$^{-1}$ corresponds to the evolution of the inter-lamellar structure factor peak. As illustrated in Figure 2 (a), the multilamellar stack consists of lipid bilayers separated by water layers.[24] The analysis of our SANS data reveals the formation of multilamellar vesicles. The best fits are obtained for $N = 2 \pm 1$ layers (0 mM) $N = 3 \pm 1$ layers (150 mM) and $N = 4 \pm 1$ (470 mM). We can obtain a convincing agreement with the SANS data for 0 mM NaCl concentration using $N = 1$. Modeling our data required to introduce less than 10 % polydispersity for the shell thickness, $t_s$ , using a Gaussian distribution. Whereas, we have to use ~ 30% polydispersity to model the inter-bilayer water thickness, $t_w$, using a Gaussian distribution. This is plausible, because some of the water is located in the polar head group region, which causes a smearing over a finite range that results in an apparently higher polydispersity.[40, 41] The polydispersity of the individual vesicles were estimated by Schulz distribution that ranges from 30 to 50 %.

We used SAXS to obtain more information on the bilayers of DOPC. Figure 2 (b) illustrates the coherent scattering cross section, as a function of the momentum transfer, $Q$, at 0, 150, and 470 mM NaCl concentration. For the sake of a better comparison, the intensity values are multiplied by an arbitrary factor, *c*. For comparison, cryo-TEM experiments are more sensitive to phosphorus, thus to the phosphocholine in the head group of the lipids as illustrated in Figure 2 (c).

We calculated the ratio of the diffraction peak positions, $Q_1:Q_2:Q_3$, in Figure 2 (b) that yields 1:2:3, indicating a regular stacking of lipid bilayers for 0 and 150 mM as observed for lamellar structure.[42] However, at the highest NaCl concentration of 470 mM there is a deviation from the regular lamellar structure. This is attributed to mixed structures as seen from cryo-TEM images. Additional Cryo-TEM images are in the supplementary information. The average lamellar repeat distance, $d = (d_1 + d_2)/2$, can be deduced from the first and second order diffraction peak, with $d_i = \frac{2\pi i}{Q_i}$ for $i = 1$ and 2, respectively. For pure DOPC lipids in $D_2O$ (0 mM salt), the repeat distance, $d = 66 \pm 4$ Å, which agrees to the value of 63.1 ± 0.3 Å reported in the literature for DOPC within the experimental accuracy.[43] The size of the head group appears to be constant with an average value of $\delta_H = 7 \pm 1$ Å.

Within the framework of the Caille model introduced in the theoretical section, the solid lines represent the fits for a lamellar structure of the bilayers. To describe the experimental data a two-step analysis seems to yield the most reliable fits. First, the Caille parameter, $\eta_{\mathrm{cp}}$, was determined for different lamellar spacing, $d_1$, by fitting the first order diffraction peak around 0.1 Å$^{-1}$, keeping, *P(Q)* = 1. In a second step, the value of $\eta_{\mathrm{cp}}$ is fixed and the parameters associated with the thickness of lamellae and interlayer spacings have been determined. The fits can describe the experimental data for $N = 3 \pm 1$ layers, for 0 and 150 mM NaCl concentrations, and $N = 4 \pm 1$ layers at the highest NaCl concentration (470 mM). The increasing number of layers with increase in NaCl concentration was also observed from cryo-TEM images (Figure 2 (c)).

***Table 1*** reports a decrease in the lamellar repeat distance, *d*, but an increase in head to head bilayer thickness, $\delta_{HH}$, with increasing the NaCl concentration. Addition of divalent cations (salt) is known to affect the electrostatic interaction between the lipid bilayers.[44, 45] A similar decrease in *d* with increasing $CaCl_2$ concentration was reported by Yamada *et al.*[45] and an increase in bilayer thickness $\delta_{HH}$, by Inoko *et al.*[44] With increasing NaCl concentration the $\mathrm{Cl}^-$ ions binds to the trimethylammonium group of the lipid head. The corresponding electrostatic repulsion overcomes the Van der Waals attraction between the lipids, instigating an increase in $\delta_{HH}$. At very high salt concentration further binding of $\mathrm{Cl}^-$ ions are screened by the existing ions and electrostatic repulsion is shielded which enables the bilayer thickness to drop below its maximum.[45] We observed that for our samples an increase in $\delta_{HH}$ from 45.40 ± 0.06 to 56.00 ± 2.02 Å (at 150 mM) and then a decrease down to 47.8 ± 0.10 Å (470 mM), cf. (*Table* 1).

In our next step, we address the question whether the bending modulus changes by adding salt. Neutron spin echo spectroscopy (NSE) is a very useful tool to obtain independent information on the bending modulus.

*Table 1. Summary of parameters obtained from the modeling of SAXS and SANS experiments shown in Figure 2. The average lamellar spacing, d, size of the head, $\delta_H$, the head to head bilayer thickness, $\delta_{HH}$, the Caille parameter, $\eta_{cp}$. From SANS, we have the thickness of the individual shells, $t_s$, the perimeter radius, $R_{SANS}$, and an estimated number of lipids per vesicle or the aggregation number is given by $N_{agg}$.*

| | SAXS | | | | SANS | | |
|---|---|---|---|---|---|---|---|
| NaCl Concen-tration (mM) | $d$ (Å) | $\delta_H$ (Å) | $\delta_{HH}$ (Å) | $\eta_{cp}$ | $R_{SANS}$ (Å) | $t_s$ (Å) | $N_{agg}$ ×$10^5$ |
| 0 | 66 ± 4 | 9.00 ± 0.01 | 45.40 ± 0.06 | 0.18 ± 0.01 | 559 ± 20 | 36 ± 1 | 1.0 ± 0.05 |
| 150 | 65 ± 1 | 9.16 ± 0.01 | 56.00 ± 2.00 | 0.20 ± 0.01 | 352 ± 52 | 35± 2 | 1.2 ± 0.06 |
| 470 | 55 ± 1 | 9.15 ± 0.02 | 47.80 ± 0.10 | 0.30 ± 0.01 | 365 ± 55 | 35 ± 2 | 1.3 ± 0.06 |

## 5.2 Dynamics by Neutron Spin Echo (NSE) Spectroscopy

Figure 3 illustrates the dynamic structure factor, $S(Q,t)$, measured by NSE at three NaCl concentrations of 0, 150 and 470 mM, covering a $Q$-range from 0.04 to 0.164 Å$^{-1}$. The solid lines represent the model description using the ZG model as described in section 4. The data emphasizes deviations of the fit utilizing the ZG model which assumes height-height correlations for membrane undulation but neglects other processes. It is illustrated in the zoomed in inset of Figure 3. We obtain bending rigidities, $\kappa_\eta/(k_BT)$ = 20 ± 2, (0 mM), 30 ± 4, (150 mM), and 40 ± 5 (470 mM). Especially the bending rigidity for the 0 mM sample is a well-established value.[27] These deviations in the fit become stronger at lower Fourier times, and also seem to depend on the salt concentration. In Table 2 we report $D_t$ from DLS and the corresponding viscosities for different NaCl concentrations. The results from the ZG analysis (equation 11) are reported in Table 3 for four different cases, where the effect of solvent viscosity, $\eta = \eta_{\mathrm{D2O}}$, (cases 1 and 2) and effective solution viscosity, $\eta = \eta_{\mathrm{D2O+vesicle}}$, (cases 3 and 4) are explored.

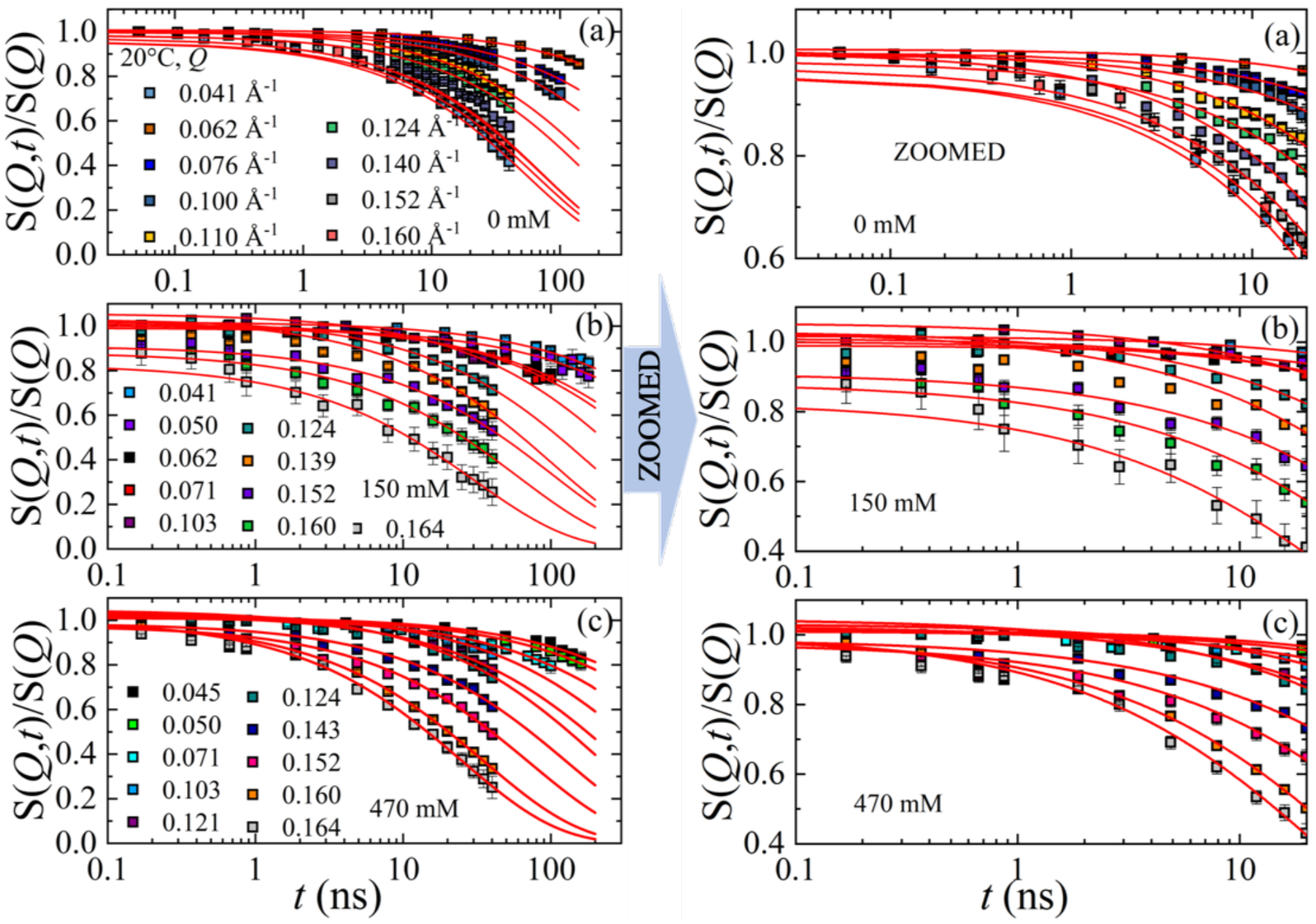

*Figure 3. Dynamic structure factor, S(Q,t)/S(Q), as a function of Fourier time, t, for different Q's for DOPC in aqueous solutions with (a) 0, (b) 150, and (c) 470 mM NaCl. The solid lines (—) represent the ZG model (equation 10). Zoomed in right panel.*

In order to better understand the observations, we will utilize the normalized mean-squared displacement, $\langle \Delta r(t)^2 \rangle_N$ or MSD.[27, 31] This approach can be used to distinguish different processes as explained below. In short, we reveal that all contributions to $S(Q,t)$ illustrated in Figure 3 originates from a motion that can be approximated by a term proportional to $\exp(-r_i^2\, Q^2)$, which justifies the calculations that lead to Figure 4. Utilizing the MSD is advantageous because it provides information without relying on a specific model. Often, changes in the slope or the respective power laws indicate different processes.

In Figure 4 (a) we compare the data in the intermediate region, i.e., we display $\langle\Delta r(t)^2\rangle_N = \langle\Delta r(t)^2\rangle / a_{ZN}$, where $a_{ZN}$ is the scaling factor. Below it becomes evident that a constant factor $a_{ZN}$ is obtained by modeling $S(Q,t)$ in the ZG region by a power-law, $a_{ZN}t^x$. Here, $x$ = 0.66, for all the samples. As discussed later, these differences imply different bending elasticities. Figure 4 (a) compares the results for 5% DOPC in the aqueous solutions with 0, 150 and 470 mM NaCl concentrations. Within the time window of our NSE experiment and in presence of NaCl, two time domains are indicated by different power-laws can be distinguished, a $t^{0.4\pm0.03}$ power-law for t < 8 ns. For higher Fourier times we obtain $t^{0.68\pm0.03}$ at 150 mM and $t^{0.70\pm0.01}$ at 470 mM. The change in power-law dependence at low Fourier times – from $t^{0.26\pm0.03}$ for pure DOPC to $t^{0.4\pm0.03}$ at 150 mM and 470 mM NaCl – indicates little differences for fast motions between the salt concentrations.

For the sake of the convenience of the reader we added the results on pure DOPC, published earlier.[27] In our previous work on a variety of vesicles (all with 0 mM salt concentration), we were able to distinguish three regions (i) $t^{0.26\pm0.03}$ for $t$ < 3 ns, (ii) $t^{0.66\pm0.01}$ for 3 ns < $t$ < 180 ns, and (iii) $t^{1.00\pm0.01}$ for $t$ > 180 ns. The current study does not include diffusive motion as it is outside the time range of Figure 3. It should be noted that, the results from molecular dynamics (MD) simulation for palmitoyl-oleoyl-phosphatidylethanolamine (POPE) predicts a power-law slope of $t^{0.37\pm0.01}$, for $t$ < 3 ns.[46]

We assume that the $t^{0.66}$ scaling behavior corresponds to the height-height correlations, known from the ZG model to be caused by thermal fluctuations. However, it is also compatible with the anomalous diffusion predicted by Monte Carlo simulations.[32,47] Both models can be considered to be Gaussian and well-compatible with $\alpha_2(t) = 0$. The same argumentation can be used for the translational diffusion.

The MSD for $t < 8$ ns shows a finite non-Gaussianity, $\alpha_2(t) = \frac{d}{d+2}\frac{\langle\Delta r(t)^4\rangle}{\langle\Delta r(t)^2\rangle^2} - 1$. As mentioned earlier, most of the signal is from those protons in the tails, thus $S(Q,t)$ or the related MSD mainly reflect the relaxation of the lipid tails. It seems to be related to the tail motion.[31] More precisely the number of protons in this case reflects the average scattering length density.

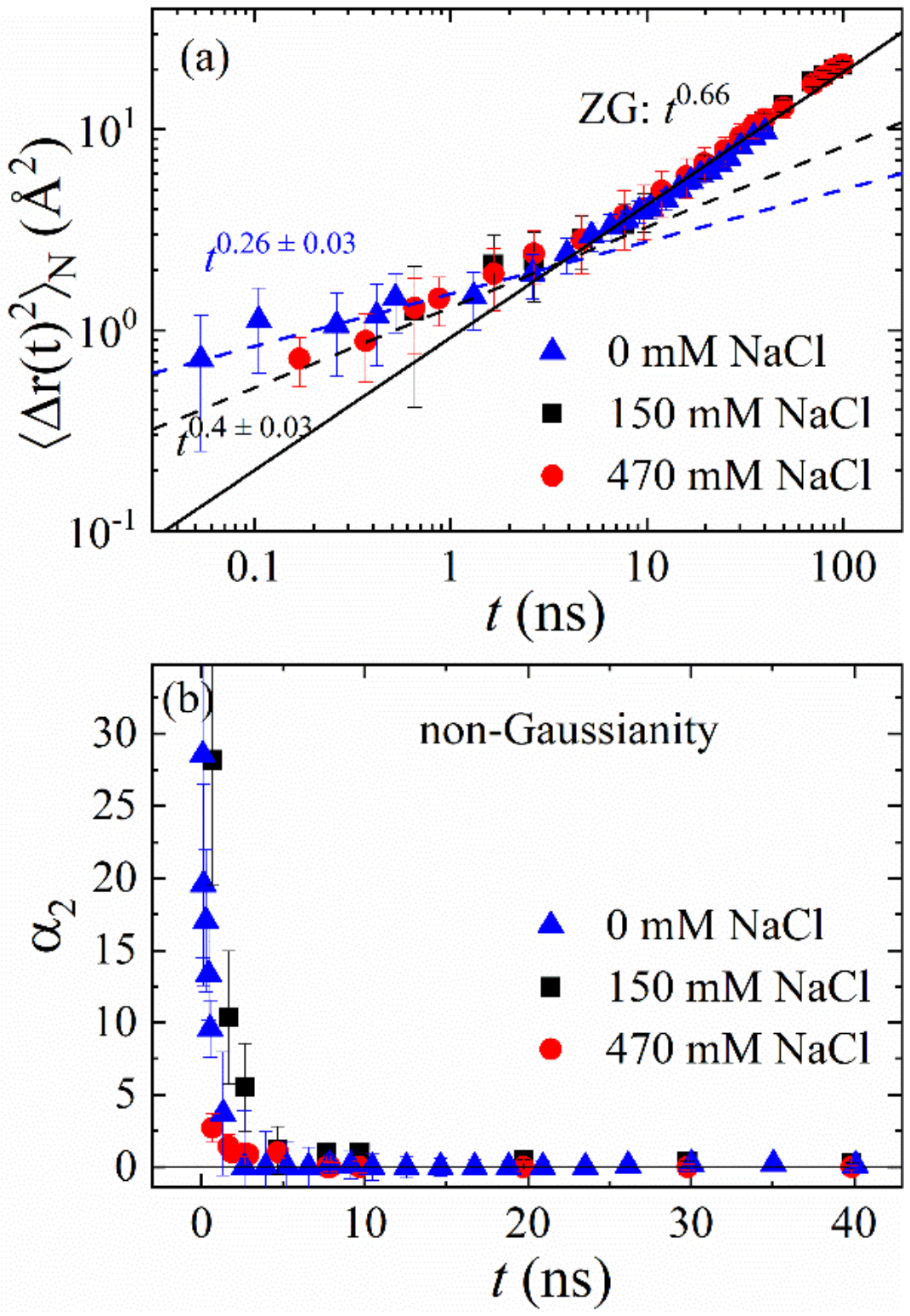

*Figure 4. (a) The normalized mean squared displacement $\langle \Delta r(t)^2 \rangle_N$ as a function of Fourier time, t, calculated from $S(Q, t)$ for 0, 150 and 470 mM NaCl, dispersed in DOPC samples. The values for pure DOPC (0 mM) were taken from our previous study.[27] The solid and dashed lines represent the experimental power-law dependence. (b) The corresponding non-Gaussian parameters, $\alpha_2(t)$, as function of t.*

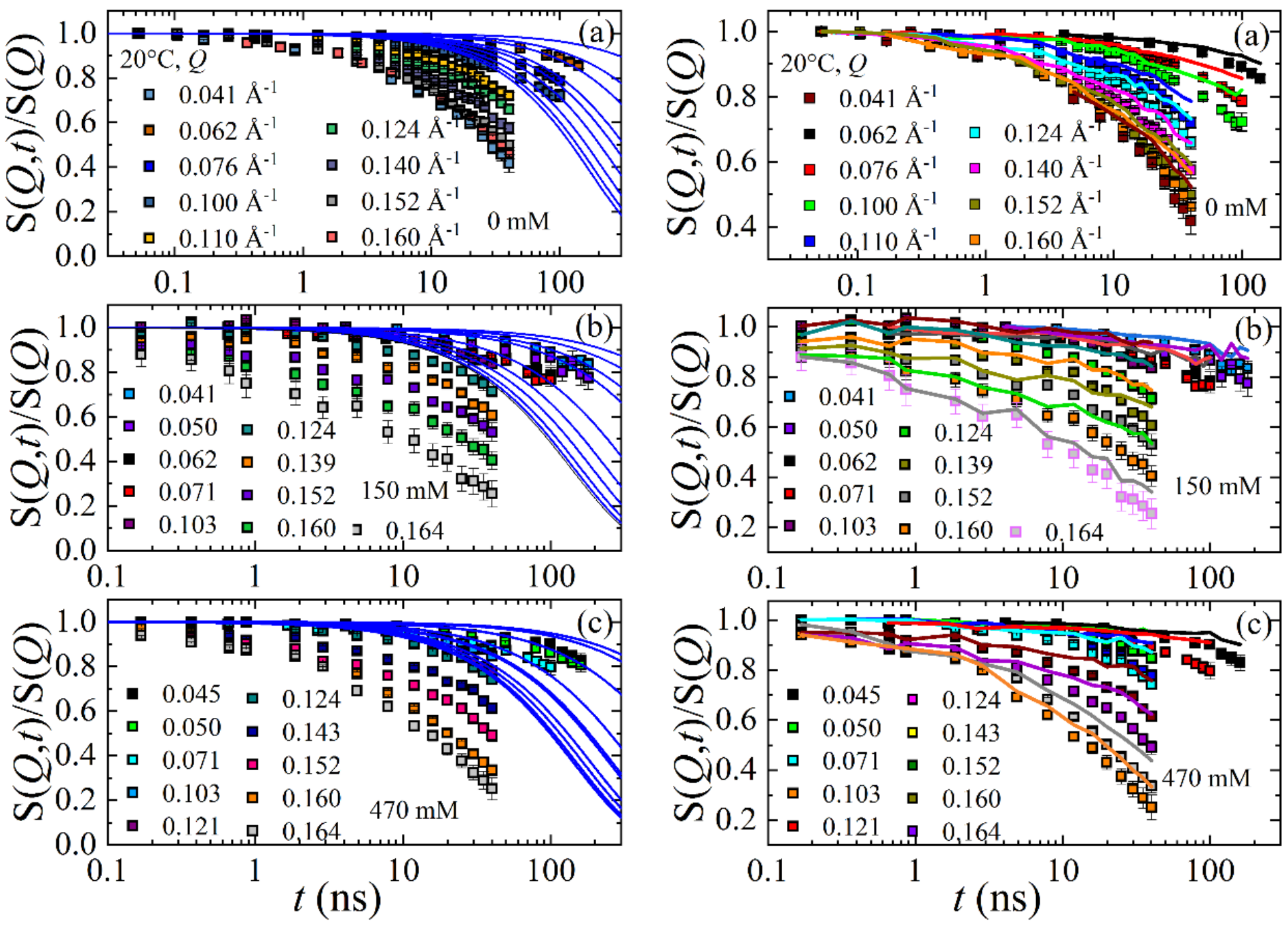

*Figure 5. Influence of the translational diffusion of the vesicles on the dynamic structure factor, S(Q,t)/S(Q), for (a) 0, (b) 150, and (c) 470 mM NaCl. The solid lines (—) in the left column represent the calculation for translational diffusion, $exp(-D_t Q^2 t)$, from low to high Q's (top to bottom). The solid lines in the right column represent experimental data divided by $exp(-D_t Q^2 t)$. It highlights the deviation in the data introduced by the diffusion.*

In Table 2 we report $D_t$ from DLS for different NaCl concentrations. In Figure 5 a plot of the three S($Q$,t) for 0, 150 and 470 mM salt data and the $S(Q,t) = \exp(-D_t Q^2 t)$, shows that there is a likely impact of the translational diffusion that needs to be considered, though the MSD does not show any effect. Specially for $t > 10$ ns, and for $Q$ > 0.07 Å $^{-1}$ the observed NSE relaxation spectra are significantly influenced by the translational diffusion of the liposomes.

Taking this experimental observation into account, we can model the intermediate scattering function, $S(Q,t)$, using the fact that we have at least three different processes, the translational diffusion of the vesicle, the ZG membrane undulation and the contribution from tail motion of the lipids to $S(Q,t)$. It can be modeled using equation 9. Figure 6 represents the corresponding data modeling for different NaCl concentrations. It can describe the experimental data including the lower Fourier times as shown in the zoomed image.

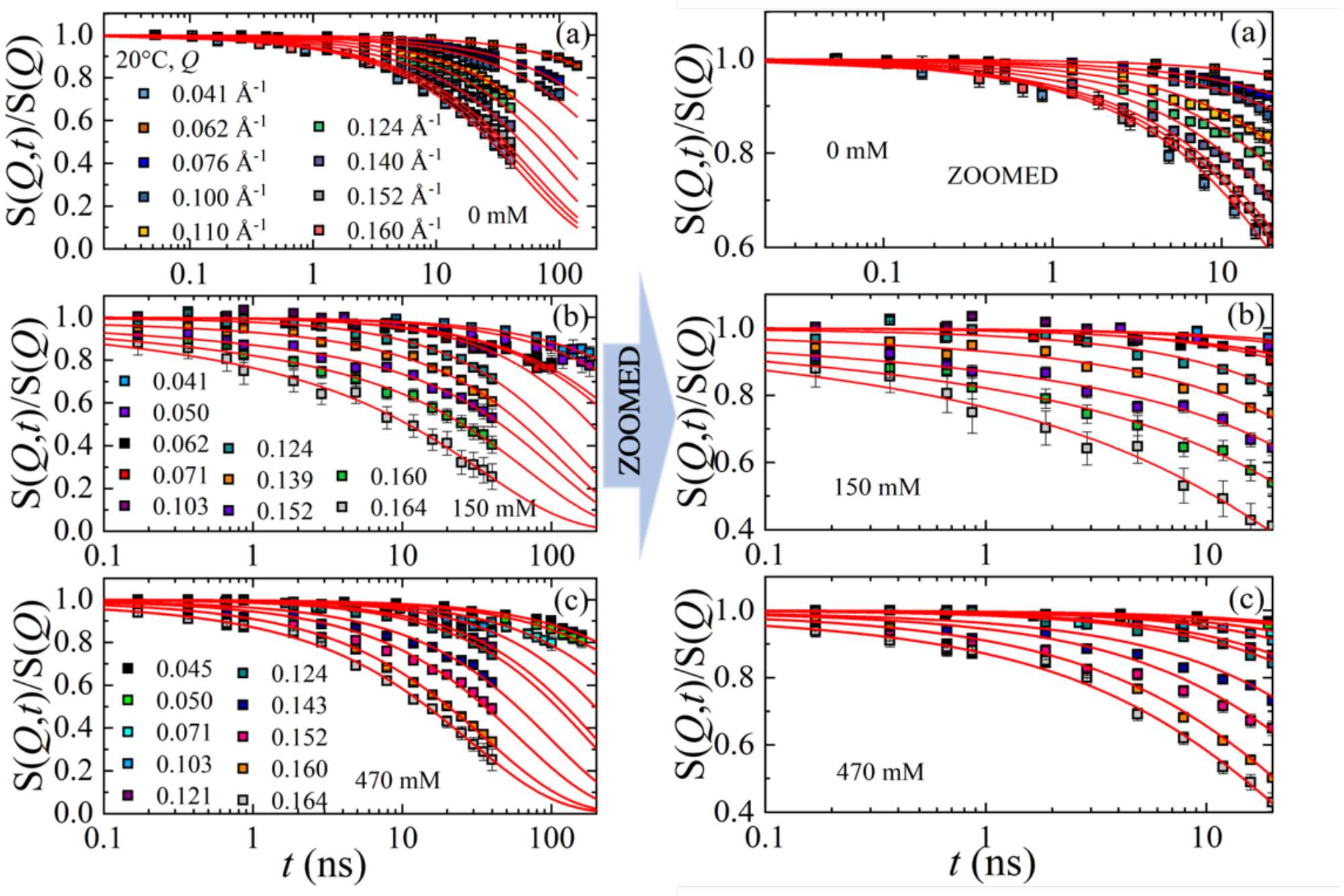

*Figure 6. Dynamic structure factor, S(Q,t)/S(Q), as a function of Fourier time, t, for different Q's as indicated for (a) 0 mM (b) 150 mM, and (c) 470 mM NaCl samples. The solid (—) lines represent the model following the discussion in equation 9. Zoomed in right panel.*

In a next step, we utilize a model free attempt to shed more light on those changes that are caused by the addition of salt. Equation 13 defines $\kappa_\eta/(k_BT)$ as a function of the Fourier time.[31] This derivation assumes the ZG height-height correlations determine the relaxation in the full-time window of our NSE from the MSD calculated in Figure 4, and thus provides an independent procedure to explore the data. However, it can also utilize the MSD, either from experiments or simulations, to simplify the result by avoiding a $Q$ dependence, thus provides an independent test to locate additional dynamics in MSD modelling. In the limit $Q \to 0$ we can use the approximation $\kappa_\eta/k_BT \propto t^{2-3x}$, for the ZG prediction, by assuming the effect of translational diffusion of the vesicle is negligible ($D_t = 0$), and for $\alpha_2 = 0$ . For $x = 0.66$, as predicted by ZG model we can determine $\kappa_\eta/k_BT$ is independent of $t$. A deviation from $t$ independent behavior is reflected by $\alpha_2 \neq 0$, or, $D_t \neq 0$, or the presence of additional dynamics.[48, 49]

Figure 7 represents the calculated membrane rigidity, $\kappa_\eta / k_B T$ as a function of the Fourier time over the entire NSE time window, for 0, 150, and 470 mM NaCl concentration. The conversion via equation 11 includes assumptions about the viscosity of the solvent/solution. Recently, it has been demonstrated that $\eta = \eta_{solvent}$, however, we need to test this result critically as salt modifies the interaction potential for our vesicles.[27, 29] We also witness the impact of a finite diffusion on $S(Q, t)$ as shown in Figure 5.

Let's consider four different cases, as presented in Table 3. The membrane rigidity, $\kappa_\eta$, is calculated with (cases 2 and 4) and without (cases 1 and 3) the translational diffusion, $D_t$, of the vesicle. The effect of solvent viscosity, $\eta = \eta_{\mathrm{D2O}}$, (cases 1 and 2) and effective solution viscosity, $\eta = \eta_{\mathrm{D2O+vesicle}}$, (cases 3 and 4) in the ZG model (equation 11) are additionally included. Having a finite value of translational diffusion (cases 2 and 4), we are also considered the effect by dividing the data by translational diffusion, $\exp(-D_t Q^2 t)$, in Figure 5. This is equivalent to subtracting the corresponding effect $(6D_t t)$ from the MSD in Figure 4. In Figure 7 the time dependence for t < 10 ns reflects the deviation from the traditional ZG model. Here the power-law dependence is a direct consequence of the tail motion power-law dependence of the MSD presented in Figure 4. We obtained, $\kappa_\eta / k_B T \propto t^{1.22}$, using $x = 0.26$ for pure DOPC and $\kappa_\eta / k_B T \propto t^{0.8}$, using $x = 0.4$ for 150 mM and 470 mM NaCl, samples. The time window $10$ ns $\leq t \leq 100$ ns (highlighted), is the range in which we expect the ZG motion with, $\kappa_\eta / k_B T \propto t^0$, for $x = 0.66$. Within the experimental uncertainty, we observed a constant value for 0 mM and 150 mM NaCl samples. However, at high $t$ for 470 mM we observed a deviation and the experimental data shows a $t^{0.5}$ dependence, which is attributed to additional dynamics apart from membrane undulation.[48]

From these calculations one can obtain the corresponding average bending rigidities, $\kappa_\eta = 32 \pm 3\ k_{\mathrm{B}}T$ for 150 mM and $34 \pm 3\ k_{\mathrm{B}}T$ for 470 mM NaCl concentration, presented by the filled symbols, case 1 in Figure 7 (a). In comparison, the bending elasticity of the 0 mM sample equals $18 \pm 2\ k_{\mathrm{B}}T$, presented by the horizontal lines. Thus, we observe an increase in membrane rigidity at higher NaCl concentrations which seems to be unchanged at 150 and 470 mM samples. However, in case 2 if we take into account the MSD from the translational diffusion of the vesicles ($D_t \neq 0$), we observe a substantial increase in the bending rigidities, $\kappa_\eta = 33 \pm 7\ k_{\mathrm{B}}T$ for 0 mM, $\kappa_\eta =$ $92 \pm 13\ k_B T$ for 150 mM and $81 \pm 12\ k_B T$ for 470 mM NaCl concentrations, presented in Figure 7 (b). Taking into consideration the solution viscosities from rheology (cf. Supplemental information) if we can recalculate, $\kappa_\eta$, for $D_t = 0$ (case 3) and $D_t \neq 0$ (case 4) and observed a relative increase in $\kappa_\eta$. The results for the average bending rigidities are reported in Table 3 for all the four different cases. In our previous work we have demonstrated that case 1 and case 4 yields more physically realistic values of $\kappa_\eta$, for pure DOPC compared to other techniques.[27]

The increase in the bending rigidity with higher NaCl concentrations as observed by NSE (case 1 and 4 in Table 3) can be explained as follows. Our SAXS indicate an increase in the bilayer thickness, $\delta_{HH}$, with increasing the NaCl concentration. It was reported in the literature that the bending rigidity increases strongly with increase in bilayer thickness.[50, 51] In our case with increasing ionic concentration the membrane dehydration can cause the lipids to densely pack, contributing to the overall increase in rigidity.

In Table 4 we have calculated the relative membrane rigidity $\widetilde{\kappa_\eta} = \kappa_\eta(\phi)/\kappa_\eta(\phi = 0)$ to compare the concentration effect of NaCl in different models. All the models indicate increase in the rigidity with respect to 0 mM NaCl, whereas the difference between 150 and 470 mM is negligible. The calculation reveals that the rigidity, $\kappa_\eta(\phi)$, from our new model and the MSD analysis is lower than that obtained using ZG model. This time-dependent analysis for the bending rigidity highlights a limited ZG time window where the rigidity is physically meaningful.

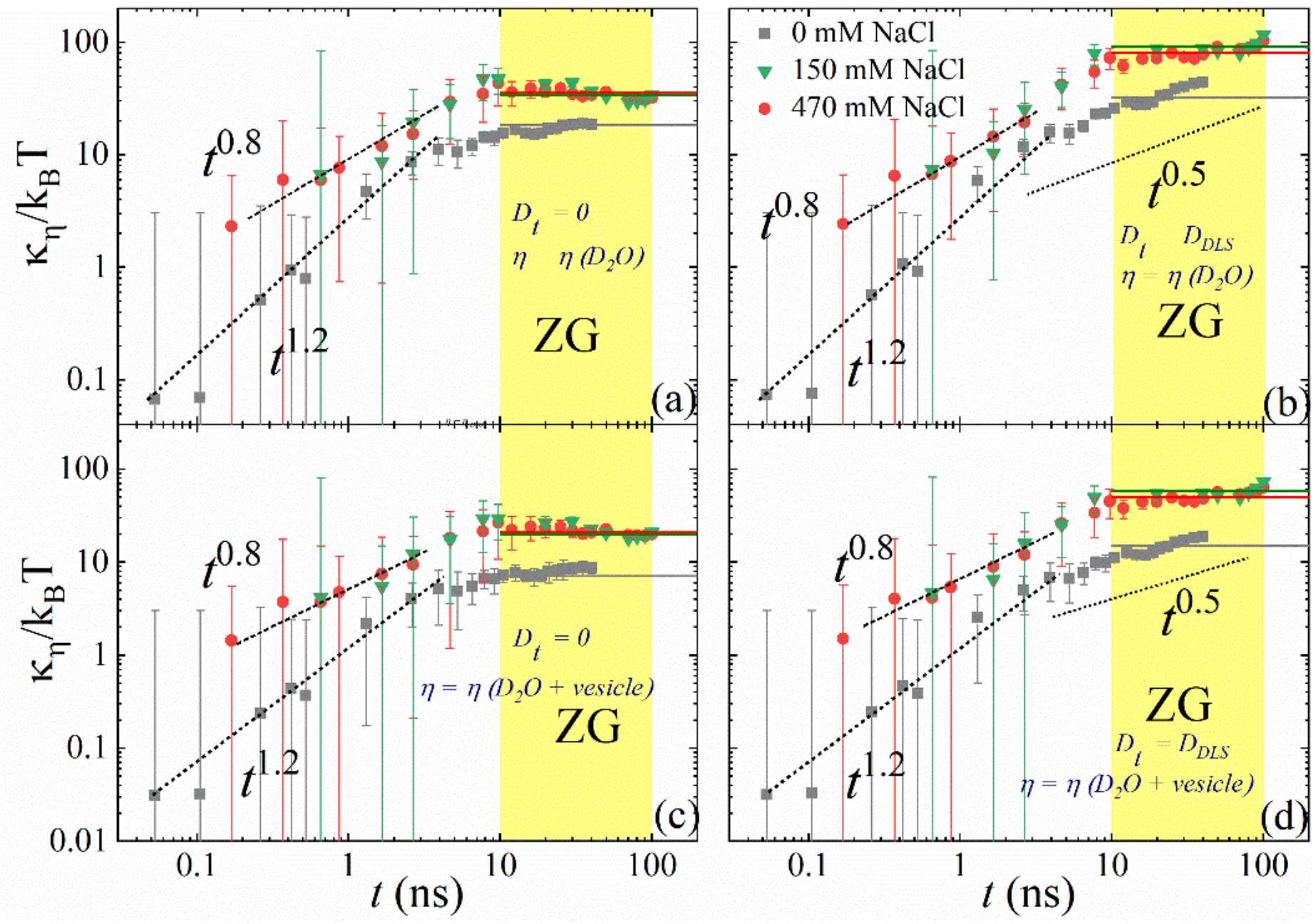

*Figure 7. The membrane rigidity, $\kappa_\eta$, divided by thermal energy, $k_BT$ , with the Boltzmann constant, $k_B$ , and the temperature, $T$, as a function of Fourier time. The data is calculated over the NSE time window from the MSD* in Figure 4 *, for 0, 150 and 470 mM NaCl concentrations, for the cases: (1) $D_t = 0$, $\eta = \eta_{D2O}$, (2) $D_t = D_{DLS}$, $\eta = \eta_{D2O}$, (3) $D_t = 0$, $\eta = \eta_{D2O+vesicle}$, and (4) $D_t = D_{DLS}$, $\eta = \eta_{D2O+vesicle}$. The parameters can be found Table 2. The yellow area indicates the time range in which MSD $\propto t^{0.66}$. We calculated the average values in this region and included this as horizontal lines for each data set. These lines represent the bending modulus, $\kappa/k_BT$ and the values are listed in Table 3. The different power-laws are explained in the text.*

*Table 2. $D_{DLS}$: diffusion coefficient from DLS; $\eta_{D2O+vesicle}$: solution viscosity in presence of NaCl, from micro-viscometry. Here the errors representing one standard deviation.*

| Concentration NaCl (MM) | 0 | 150 | 470 |
|---|---|---|---|
| $D_{DLS}$ (Å$^2$ns$^{-1}$), at 5 wt% | 0.22 ± 0.01 | 0.27 ± 0.01 | 0.26 ± 0.01 |
| $\eta_{D2O+VESICLE}$ (mPa·s), at 5 wt% | 1.91 ± 0.04 | 1.58 ± 0.03 | 1.59 ± 0.03 |

*Table 3. Membrane rigidity $\kappa_\eta$ obtained using different models. We distinguish four cases: (1) $D_t = 0$, $\eta = \eta_{D2O}$, (2) $D_t = D_{DLS}$, $\eta = \eta_{D2O}$, (3) $D_t = 0$, $\eta = \eta_{D2O+vesicle}$, and (4) $D_t = D_{DLS}$, $\eta = \eta_{D2O+vesicle}$. We use 5wt% DOPC lipid concentration in $D_2O$.*

| | | $\kappa_\eta/(k_B T)$ | | | |
|---|---|---|---|---|---|
| **Model description** | **Concentration NaCl (mM)** | **Case 1:** $D_t = 0$, $\eta = \eta_{D2O}$ | **Case 2:** $D_t = D_{DLS}$, $\eta = \eta_{D2O}$ | **Case 3:** $D_t = 0$, $\eta = \eta_{D2O+vesicle}$ | **Case 4:** $D_t = D_{DLS}$, $\eta = \eta_{D2O+vesicle}$ |
| ***ZG analysis and Diffusion (Full time range)*** | 0 | 26 ± 1 | 53 ± 6 | 11 ± 1 | 23 ± 2 |
| | 150 | 34 ± 4 | 118 ± 30 | 21 ± 3 | 74 ± 19 |
| | 470 | 36 ± 3 | 136 ± 15 | 22 ± 2 | 84 ± 9 |
| ***ZG analysis (t >5 ns) (figure 3)*** | 0 | 20 ± 2 | 44 ± 3 | 9 ± 1 | 19 ± 2 |
| | 150 | 33 ± 9 | 89 ± 16 | 21 ± 6 | 56 ± 10 |
| | 470 | 30 ± 6 | 94 ± 15 | 18 ± 4 | 58 ± 9 |
| ***new model, equation 9 (figure 6)*** | 0 | 21 ± 2 | 59 ± 5 | 9 ± 1 | 25 ± 2 |
| | 150 | 33 ± 6 | 86 ± 11 | 20 ± 4 | 54 ± 7 |
| | 470 | 32 ± 2 | 72 ± 8 | 19 ± 2 | 44 ± 5 |
| ***MSD analysis (figure 7)*** | 0 | 18 ± 2 | 33 ± 7 | 8 ± 1 | 15 ± 3 |
| | 150 | 32 ± 3 | 92 ± 13 | 20 ± 2 | 58 ± 8 |
| | 470 | 34 ± 3 | 81 ± 12 | 21 ± 2 | 50 ± 7 |

*Table 4. Relative Membrane rigidity $\widetilde{\kappa_\eta} = \kappa_\eta(\phi)/\kappa_\eta(\phi = 0)$ obtained using different models. We distinguish four cases: (1) $D_t = 0$, $\eta = \eta_{D2O}$, (2) $D_t = D_{DLS}$, $\eta = \eta_{D2O}$, (3) $D_t = 0$, $\eta = \eta_{D2O+vesicle}$, and (4) $D_t = D_{DLS}$, $\eta = \eta_{D2O+vesicle}$. We use 5wt% DOPC lipid concentration in $D_2O$*

| | | $\boldsymbol{\kappa_\eta(\Phi)/\kappa_\eta(\phi = 0)}$ | | | |
|---|---|---|---|---|---|
| **Model description** | **Concentration NaCl (mM)** | **Case 1:** $D_t = 0$ $\eta = \eta_{D2O}$ | **Case 2:** $D_t = D_{DLS}$ $\eta = \eta_{D2O}$ | **Case 3:** $D_t = 0$, $\eta = \eta_{D2O+vesicle}$ | **Case 4:** $D_t = D_{DLS}$, $\eta = \eta_{D2O+vesicle}$ |
| ***ZG analysis and Diffusion (Full time range)*** | 150 | 1.31 ± 0.12 | 2.23 ± 0.28 | 1.91 ± 0.17 | 3.22 ± 0.27 |
| | 470 | 1.38 ± 0.09 | 2.57 ± 0.16 | 2.00 ± 0.13 | 3.65 ± 0.15 |
| ***ZG analysis (t >5 ns) (figure 3)*** | 150 | 1.65 ± 0.29 | 2.02 ± 0.23 | 2.33 ± 0.31 | 2.95 ± 0.21 |
| | 470 | 1.50 ± 0.22 | 2.14 ± 0.21 | 2.00 ± 0.25 | 3.05 ± 0.19 |
| ***new model, equation 9 (figure 6)*** | 150 | 1.57 ± 0.21 | 1.46 ± 0.21 | 2.22 ± 0.23 | 2.16 ± 0.15 |
| | 470 | 1.52 ± 0.11 | 1.22 ± 0.20 | 2.11 ± 0.15 | 1.76 ± 0.14 |
| ***MSD analysis (figure 7)*** | 150 | 1.78 ± 0.15 | 2.79 ± 0.25 | 2.50 ± 0.16 | 3.87 ± 0.24 |
| | 470 | 1.89 ± 0.14 | 2.45 ± 0.26 | 2.63 ± 0.16 | 3.33 ± 0.24 |

## 6 CONCLUSION

We have studied structural and dynamic changes in zwitterionic DOPC liposomes at the two important NaCl concentrations 150 mM, relevant to physiological concentration and 470 mM, relevant to the average oceanic concentration. We have employed state-of-the-art NSE spectroscopy to investigate the membrane dynamics in the lipid bilayer. We reported that even for, $t > 10$ ns, and for, $Q > 0.07$ Å $^{-1}$, the NSE relaxation spectra is influenced by the translational diffusion coefficient of the liposomes. In our new model considering the tail motion of the lipid we obtain an increase in bending rigidity at higher salt concentrations with and without finite translational diffusion of the vesicle, $D_t$. The increase in bending rigidity can be explained on the basis of an increased bilayer thickness due to membrane dehydration observed from SAXS. The inclusion of $D_t$ in the simple ZG model over the full NSE time window results in an unphysical increase in bending rigidity. However, restricting the ZG model over a limited time window, $5 < t \leq 100$ ns, yields a bending rigidity that is comparable to our new model. We introduced a time-dependent analysis for the bending rigidity that highlights only a limited ZG time window where the rigidity is constant and physically meaningful. The measured data shows deviations from that model which indicates onset of secondary dynamics, which can be further explored in the future. The evident increase in bending rigidity of vesicles with zwitterionic head group and unsaturated hydrocarbon chains as a result of external NaCl implies that this effect has to take into consideration in liposomal drug formulations and other applications. The DOPC vesicles are studied far from the transitioning temperature $T_m$, which is one of the prominent factors affecting bending rigidity as the system lipids cross over from $L_o$ to $L_d$, and we demonstrate that subtle changes in external salt concentration can drive changes in collective motions such as membrane undulations and thickness changes which increases the rigidity in the fluid phase lipid vesicles.

## AUTHOR INFORMATION

**Corresponding Authors**

*E-mail: g.sudipta26@gmail.com

*E-mail: gjschneider@lsu.edu

**ORCID**

Judith U. De Mel: 0000-0001-7546-1491

Sudipta Gupta: 0000-0001-6642-3776

Gerald J. Schneider: 0000-0002-5577-9328

**Notes**

The authors declare no competing financial interest.

## ACKNOWLEDGEMENT

The neutron scattering work is supported by the U.S. Department of Energy (DOE) under EPSCoR Grant No. DE-SC0012432 with additional support from the Louisiana Board of Regents. Access to the neutron spin echo spectrometer and small-angle scattering instruments was provided by the Center for High Resolution Neutron Scattering, a partnership between the National Institute of Standards and Technology and the National Science Foundation under Agreement No. DMR-1508249. Research conducted at the Spallation Neutron Source (SNS) at Oak Ridge National Laboratory (ORNL) was sponsored by the Scientific User Facilities Division, Office of Basic Energy Sciences, U.S. DOE. We thank Lin Yang and Shirish Chodankar from 16-ID, LIX beamline at National Synchrotron Light Source (NSLS) II. The LiX beamline is part of the Life Science Biomedical Technology Research resource, primarily supported by the National Institute of Health, National Institute of General Medical Sciences under Grant P41 GM111244, and by the Department of Energy Office of Biological and Environmental Research under Grant KP1605010, with additional support from NIH Grant S10 OD012331. As a NSLS II facility resource at Brookhaven National Laboratory, work performed at Life Science and Biomedical Technology Research is supported in part by the US Department of Energy, Office of Basic Energy Sciences Program under Contract DE-SC0012704. CG3 Bio-SANS instrument and HFIR are sponsored by the Office of Biological and Environmental Research, and the Scientific User Facilities Division, Office of Basic Energy Sciences, U.S. Department of Energy, respectively. We would like to thank Rafael Cueto (LSU) for his support in DLS experiments, Ralf Biehl (Forschungszentrum Jülich GmbH, Germany) for his support in performing viscosity measurements and Jibao He (Tulane University, USA) for his support in Cryo-TEM. We would like to acknowledge Antonio Faraone (NIST) for his support to the NSE experiments and Cedric Gagnon (NIST) for his support at the VSANS beamline.

## DISCLAIMER

Certain trade names and company products are identified in order to specify adequately the experimental procedure. In no case does such identification imply recommendation or endorsement by the National Institute of Standards and Technology, nor does it imply that the products are necessarily the best for the purpose. This paper was prepared as an account of work sponsored by an agency of the United States Government. Neither the United States Government nor any agency thereof, nor any of their employees, makes any warranty, express or implied, or assumes any legal liability or responsibility for the accuracy, completeness, or usefulness of any information, apparatus, product, or process disclosed, or represents that its use would not infringe privately owned rights. Reference herein to any specific commercial product, process, or service by trade name, trademark, manufacturer, or otherwise does not necessarily constitute or imply its endorsement, recommendation, or favoring by the United States Government or any agency thereof. The views and opinions of authors expressed herein do not necessarily state or reflect those of the United States Government or any agency thereof.